\documentclass[twocolumn,showpacs,preprintnumbers,amsmath,amssymb]{revtex4}
\usepackage{dcolumn}
\usepackage{bm}
\usepackage{graphicx}
\usepackage{amsmath}
\begin{document}
\title{Spatiotemporal Bloch states of a spin-orbit coupled Bose-Einstein condensate in an optical lattice\footnote{Project supported by the National Natural Science Foundation of China (Grant No. 11475060).}}
\author{Ya-Wen Wei, \ Chao Kong, \ Wen-Hua Hai\footnote{Corresponding author. Email address:
whhai2005@aliyun.com}}
\affiliation{Department of Physics and Key Laboratory of Low-dimensional Quantum Structures and Quantum Control of Ministry
of Education, and Synergetic Innovation Center for Quantum Effects and Applications, Hunan Normal University, Changsha 410081, China}

\begin{abstract}
We study the spatiotemporal Bloch states of a high-frequency driven two-component Bose-Einstein condensate (BEC) with spin-orbit coupling (SOC) in an optical lattice. By adopting the rotating-wave approximation (RWA) and applying an exact trial-solution to the corresponding quasistationary system, we establish a different method for tuning SOC via external field such that the existence conditions of the exact particular solutions are fitted. Several novel features related to the exact states are demonstrated, such as SOC leads to spin-motion entanglement for the spatiotemporal Bloch states, SOC increases the population imbalance of the two-component BEC and SOC can be applied to manipulate the stable atomic flow which is conducive to control quantum transport of the BEC for different application purposes.

\end{abstract}


\maketitle

\section{Introduction}

SOC, the interaction between the spin and momentum of a quantum particle, is crucial for many important condensed matter phenomena \cite{1}. As is known to all, due to the groundbreaking work of Dresselhaus and Rashba and the further theories and experiments they initiated \cite{2,3,4,5,6,7}, the physical relevance of SOC is found. The results were extended to the ultracold atomic systems, where various synthetic SOC can be induced and managed by external laser field. In particular, SOC has been experimentally realized \cite{8} and theoretically investigated \cite{9,10,11} for binary mixtures of BEC. The cold atomic systems confined in an optical lattice with SOC have also been widely studied. In the mean-field theory, as a many-body system, a BEC governed by Gross-Pitaevskii equation (GPE) provides an important basis for studying the corresponding physical properties.
It is important to find exact solutions \cite{12,13} and analytically perturbed solutions \cite{14,15} to the GPE, which can be used to discuss many physical properties such as the macroscopic quantum (or semi-classical) chaos \cite{14,15,16,17}, BEC stability \cite{18,19,20,21,22}, the superfluid velocity and flow density \cite{23}, and the generation of solitons \cite{21}. At present, the researchers have obtained some exact stationary solutions to the BEC system in the quasi-one-dimensional (quasi-1D) Kronig-Penny potential \cite{24} and optical lattice potential \cite{19,20,25,26}, and in the two-dimensional optical lattice potential \cite{27,28}. However, the GPE with nonlinear interactions is not easy to solve in general, since the exact solution can exist only for a certain parameter conditions which are hard to find. The BEC system with SOC \cite{29,30} is a more complex nonlinear system which poses a challenge to finding the exact solutions. The high frequency approximation method has been widely used, since the corresponding approximate systems can be some exactly solvable quasistationary ones \cite{31,32}. Doing so is called the RWA for
historical reasons \cite{31}. Recently, by means of the high frequency approximation, the SOC tunability has been investigated theoretically and experimentally \cite{33,34}. The SOC strength, of course, can also be adjusted by the magnitude and direction of the Raman laser wave-vector. Here we are motivated in generation of the exact solutions to such a quasistationary system with SOC, by using a high-frequency external field to adjust and reconstruct the system parameters. Consequently, it is possible to control physical properties of the system and to make quantum transitions between the different analytical quantum states.

The spin-motion entangled state of a single atom has been defined well and was employed in many previous works \cite{35,36,37}, which contains the well-known ¡°Schr\"odinger cat¡± state \cite{38} $|\Psi(t)\rangle=\frac{1}{\sqrt{2}}(|x_{1}\rangle|\uparrow\rangle+|x_{2}\rangle|\downarrow\rangle)$ with the spin-up and spin-down internal states $|\uparrow\rangle$ and $|\downarrow\rangle$ and the coherent motional states $|x_{i}\rangle$. For a SO coupled BEC trapped in a superlattice, the similar spin-motion entangled states can be precisely defined by using two linearly independent motional states to replace the coherent motional states $\frac{1}{\sqrt{2}}|x_{i}\rangle$ \cite{39}. In the full-quantum treatment, these spin-motion entangled states may be implemented to encode the qubits \cite{35,36,37,40}, and the quantum information can be transported through the spin-dependent currents \cite{41,42}. Therefore, we are also interested in how to generate the spin-motion entanglement and to manipulate the spin-dependent currents by tuning the SOC strength and other parameters \cite{33,34}.

In this paper, we consider a SO-coupled BEC driven by a high-frequency field and confined in an optical lattice through which we study the spatiotemporal Bloch states \cite{43} and the associating physical properties. First, based on the RWA, such a driven system is approximated by a quasistationary one and the relating exact spatiotemporal Bloch states to the latter system are obtained. We analytically show that the SOC strength can be adjusted by the high-frequency field such that the existence conditions of the exact particular solutions are met. Meanwhile we find that areas of the parameter regions on the SOC- vs lattice-strength plane for the existence of the exact solution enlarge with the decrease in Rabi coupling strength. Then, we show the periodic distributions of the atomic number density, which contains the discrete zero points, corresponding to the instability \cite{22}.
Furthermore, we analytically and numerically demonstrate the several novel features related to the exact states: (a) the SOC leads to that the spatiotemporal Bloch states become the spin-motion entangled states; (b) the SOC influences the population imbalance between two BEC components and may result in that all the atoms completely occupy one of the hyperfine states; (c) SOC can be applied to manipulate the stable atomic flow, which is conducive to avoid instability and to control quantum transport of the BEC for different application purposes \cite{44,45}.

\section{Exact solutions to the quasi-stationary system}

We consider a BEC confined in a quasi-1D optical lattice potential $V(x)$ oriented in the longitudinal $x$ direction, and the transverse dynamics of the condensate is assumed to be frozen to the respective ground states of the harmonic traps. In the case of high-frequency driving, the macroscopic quantum state of the BEC is $|\Psi(x,t)\rangle=\Psi_1(x,t)|\uparrow\rangle+\Psi_2(x,t)|\downarrow\rangle$ with the spin-up and spin-down internal states $|\uparrow\rangle$ and $|\downarrow\rangle$ and the macroscopic wave function $\Psi_{j}(x,t)=\langle x|\Psi_j(t)\rangle$ which is governed by the nonlinear GP equations \cite{46}
\begin{eqnarray}\label{dy3}
i\frac{\partial\Psi_j}{\partial t}=\Big[-\frac{1}{2}\frac{\partial^2}{\partial x^2}+V(x)\Big]\Psi_j+i(-1)^j\gamma \frac{\partial\Psi_j}{\partial x}+\Gamma\Psi_{3-j}\nonumber\\
+(g|\Psi_j|^2+g_{12}|\Psi_{3-j}|^2)\Psi_j+\xi x\sin(\omega t)\Psi_j\ \text{for}\  j=1, 2\
\end{eqnarray}
with the optical lattice potential $V(x)=V_{0}\sin^{2} x$. All the terms in Eq. (1) are the same as the corresponding terms in Ref. \cite{46}, except for the driving term. The SOC and Rabi coupling have been realized experimentally \cite{8}, and the driving can be realized experimentally by using a periodic magnetic field gradient applied along the $x$ direction \cite{47}. However, the system of units we adopted is different. Here the driving frequency $\omega$ and energies (including the lattice depth $V_0$ and Rabi-coupling strength $\Gamma$) have been normalized in units of the recoil frequency $\omega_{r}=\frac{\hbar k^{2}}{m}$ and recoil energy $E_r=\hbar \omega_r$, with $m$ and $k$ being the atomic mass and standing wave vector. The spatial coordinate $x$, time $t$, and atomic number density $|\Psi_{j}|^{2}$ have also been normalized in units of $k^{-1}$, $\omega_r^{-1}$ and $k$. The lattice tilt $\xi$ and SOC strength $\gamma>0$ are in units of $kE_r$ and $E_r/k$, respectively, and the constants $g$ and $g_{12}$ are the corresponding dimensionless intra- and inter-species interaction strengths which can be adjusted independently by the optical and magnetic Feshbach-resonance techniques \cite{48} in actual experiments. By the high-frequency limit and strong field we mean that the driving frequency and strength obey $\omega\gg 1$ and $\xi/\omega \sim 1$.
Due to the particle-number conservation, we let the average total number $N_{t}=N_{1}+N_{2}$ of particles in each well be a constant with $N_{j}$ being the average atomic number per well for the $j$th component. To make use of the advantage of RWA, we seek the wave functions in the form $\Psi_{j}(x,t)=\Phi_{j}(x,t)e^{-i2\frac{\xi}{\omega}x\sin^{2}(\frac{\omega t}{2})}$ with $\Phi_{j}(x,t)$ being treated as a slowly varying function of time \cite{31,32}. By inserting such wave functions into Eq. (1) yields the equations
\begin{eqnarray}
&&i\frac{\partial\Phi_{j}(x,t)}{\partial t}=-\frac{1}{2}\frac{\partial^{2}\Phi_{j}(x,t)}{\partial x^{2}}+i\Big[\gamma(-1)^{j}+\frac{\xi}{\omega}\Big]\frac{\partial}{\partial x}\Phi_{j}(x,t)\nonumber\\
&&+V(x)\Phi_{j}(x)+\Gamma\Phi_{3-j}(x,t)+(g|\Phi_{j}|^2+g_{12}|\Phi_{3-j}|^2)\Phi_{j}(x,t)\nonumber\\
&&+\Big[\frac{\gamma(-1)^{j}\xi}{\omega}+\frac{3\xi^{2}}{4\omega^{2}}\Big]\Phi_{j}(x,t)
\end{eqnarray}
without driving, where we have replaced the fast-varying functions $\sin^{2}(\frac{\omega t}{2})$ and $\sin^{4}(\frac{\omega t}{2})$ by their average values $\frac{1}{2}$ and $\frac{3}{8}$, respectively, according to the RWA. By substituting the stationary solutions
\begin{eqnarray}
\Phi_{j}(x,t)=\psi_{j}(x)e^{-i\mu t}
\end{eqnarray}
with chemical potential $\mu$ into Eq. (2), we obtain the time-independent equations of $\psi_{j}(x)$
\begin{eqnarray}
\mu_{j}\psi_{j}&=&-\frac{1}{2}\psi_{j,xx}+i\gamma_{j}\psi_{j,x}+\Gamma\psi_{3-j}+V(x)\psi_{j}+(g|\psi_{j}|^2\nonumber\\
&&+g_{12}|\psi_{3-j}|^2)\psi_{j},\nonumber\\
\gamma_{j}&=&(-1)^{j}\gamma+\frac{\xi}{\omega},\ \
\mu_{j}=\mu-(-1)^{j}\gamma\frac{\xi}{\omega}-\frac{3\xi^{2}}{4\omega^{2}}
\end{eqnarray}
with $\gamma_j$ and $\mu_i$ being the recombined SOC strengths and chemical potentials which are tuned by the driving parameter ratio $\frac{\xi}{\omega}$.

It is worth noting that for $\gamma=0$ and $\Gamma=0$, Eq. (4) is reduced to the basic equations in Ref. \cite{25}, and the corresponding exact solutions have been constructed. The presence of SOC and Rabi coupling makes the system more complicated and more hard to solve. However, according to the famous Floquet theorem, there exists exact Bloch solution to a spatially periodic system \cite{43}.
In fact, if we adjust the driving parameter ratio $\frac{\xi}{\omega}$ and SOC strength $\gamma$ to satisfy the following conditions
\begin{eqnarray}
\mu&=&\frac{1}{2}\Big[1+N_{t}(g+g_{12})+V_{0}+\frac{3}{2}(\Gamma^{2}+\gamma^{2})\Big],\nonumber\\
\frac{\xi^{2}}{\omega^{2}}&=&\Gamma^{2}+\gamma^{2},
\end{eqnarray}
the exact particular solutions to the spatially periodic Eq. (4) can be obtained by inserting the trial Bloch solutions
\begin{eqnarray}
\psi_{j}=a_{j}\cos x+ib_{j}\sin x
\end{eqnarray}
with the real undetermined constants $a_j$ and $b_j$ into Eq. (4), and by using the normalization condition. Firstly, the normalization and Eq. (6) give the average number of particles per well in Eq. (5) as \cite{19,25} $N_t=N_1+N_2=\frac{1}{n\pi}\int_{\frac{\pi}{2}}^{(n+\frac{1}{2})\pi}\sum_{j=1}^2|\psi_{j}|^2 dx=\frac 1 2 \sum_{j=1}^2 (a_j^2+b_j^2)$. Then combining Eqs. (6) and Eq. (5) with Eq. (4) we derive the two pairs of equations $g (a_j^2-b_j^2)+g_{12}(a_{3-j}^2-b_{3-j}^2)=V_0$ and $a_j^2+b_j^2=N_t-(-1)^j\frac{2\gamma\sqrt{\Gamma^{2}+\gamma^{2}}}{g-g_{12}}$ for $j=1,2$. From the first pair we can write $a_j^2-b_j^2=a_{3-j}^2-b_{3-j}^2=\frac{V_0}{(g+g_{12})}$. Solving the four equations yields the four squares of the undetermined constants. Hereafter we take the positive constants
in the forms
\begin{eqnarray}
a_{j}&=&\sqrt{\frac{N_{t}}{2}-(-1)^j\frac{\gamma\sqrt{\Gamma^{2}+\gamma^{2}}}{g-g_{12}}+\frac{V_{0}}{2(g+g_{12})}},\nonumber\\
b_{j}&=&\sqrt{\frac{N_{t}}{2}-(-1)^j\frac{\gamma\sqrt{\Gamma^{2}+\gamma^{2}}}{g-g_{12}}-\frac{V_{0}}{2(g+g_{12})}}
\end{eqnarray}
for $j=1,2$. Making use of Eqs. (5) and (7), we can easily prove that Eq. (6) is just a pair of exact solutions to Eq. (4).
For the fixed parameters $\Gamma$ and $\gamma$, the first condition of Eq. (5) can be satisfied by the undetermined chemical potential and the second condition is adjusted by the driving parameter ratio $\frac{\xi}{\omega}$. The latter condition means that \emph{the system (4) with nonzero SOC and/or Rabi strengths has the exact solution (6), if the ratio $\frac{\xi}{\omega}$ is equal to the value of $\sqrt{\Gamma^{2}+\gamma^{2}}$}. The result is similar to the well-known important one that when the driving parameter ratio coincides with a zero of the zero order Bessel function, a high-frequency driven two-state model arrives at a particular localized state, namely the \emph{coherent destruction of tunneling} \cite{32}.

Rewriting Eq. (6) in its exponential form
\begin{eqnarray}
\psi_{j}&=&R_{j}(x)\exp[i\Theta_{j}(x)],\ \ \ \ \ j=1, 2, \nonumber\\
R_{j}&=&\sqrt{a_{j}^{2}+(b_{j}^{2}-a_{j}^{2})\sin^{2}x}, \nonumber\\
\Theta_{j}&=&\arctan \frac{b_{j}\sin x}{a_{j}\cos x}=\int\frac{a_{j}b_{j}}{R_{j}^{2}}dx,
\end{eqnarray}
and applying Eq. (7) to Eq. (8) lead to the atomic number densities
\begin{eqnarray}
R_j^{2}=\frac{1}{2}\Big[N_{t}+\frac{V_{0}\cos(2x)}{g+g_{12}}-(-1)^j\frac{2\gamma\sqrt{\Gamma^{2}+\gamma^{2}}}{g-g_{12}}\Big].
\end{eqnarray}
Clearly, the exact solutions in Eq. (8) have the similar forms with those in Ref. \cite{19,25} and in Ref. \cite{19,20,21,22} with zero elliptic modulus for the spatially periodic nonlinear systems without SOC. However, to construct such a solution to the system (4) with SOC, the high-frequency driving with the parameters obeying Eq. (5) must be applied. Given Eq. (8), we can calculate the average atomic numbers per well as \cite{19,25}
\begin{eqnarray}
N_j=\frac{1}{n\pi}\int_{\frac{\pi}{2}}^{(n+\frac{1}{2})\pi}R_{j}^{2}dx=\frac{N_{t}}{2}-(-1)^j\frac{\gamma\sqrt{\Gamma^{2}+\gamma^{2}}}{g-g_{12}}.
\end{eqnarray}
In order to keep the positive semidefinite property of the atomic number densities, $|\psi_{j}|^{2}=R^{2}_{j}\geq0$ for any $x$ value, from Eq. (9) we have the necessary and sufficient conditions \cite{25}
\begin{eqnarray}
|V_{0}|&\leq& V_{c}=\min\{V_{1c},V_{2c}\}, \nonumber\\
V_{jc}&=&|g+g_{12}|\Big[N_{t}-(-1)^j\frac{2\gamma\sqrt{\Gamma^{2}+\gamma^{2}}}{g-g_{12}}\Big].
\end{eqnarray}
The exact solutions Eqs. (6) and (7) to the quasistationary Eq. (4) are valid if and only if the depth $V_{0}$ of the lattice potential obeys the constraint relation Eq. (11). According to Eq. (10), we know that the quantity in the square brackets is greater than zero, due to $N_j\ge 0$ for any $j$, which gives a limit to the maximal value of $\gamma$ for a set of fixed other parameters. Therefore, for $g-g_{12}>0$ or $g-g_{12}<0$, Eq. (11) gives $V_c=V_{2c}$ or $V_c=V_{1c}$ respectively.
Without loss of the generality, we can suppose $V_{0}\geq 0$.
Then based on Eq. (11), taking the parameters $N_{t}=10$, $g=0.6$ and $g_{12}=0.4$, we plot the boundary curves $V_0=V_{2c}(\gamma)$ for the parameter regions of the exact solution under the RWA as a function of $\gamma$ for $\Gamma=0.1$, $0.8$, and $1.4$, respectively, as shown in Fig. 1(a).
Any curve in Fig. 1(a) with a fixed $\Gamma$ value gives a boundary corresponding to the parameter regions for the existence and nonexistence of the exact solution (6). Noticing the sign ``$\le$" in Eq. (11), the parameter regions of exact solution (6) are A, (A+B) and (A+B+C) for $\Gamma=1.4$, $0.8$ and $0.1$, respectively, which means that the stronger Rabi coupling decrease the size of the considered parameter region.
In Fig. 1(b) based on $V_0=V_{1c}(\gamma)$, we also show the existence regions of the exact solutin for $N_{t}=8$ and the different nonlinear interaction strengths $g=0.4$ and $g_{12}=0.6$. Similar to Fig. 1(a), the parameter regions for the existence of the exact solution still are A, (A+B) and (A+B+C) for $\Gamma=1.4$, $0.8$ and $0.1$, respectively. The areas of the existence regions enlarge with the decrease in Rabi coupling strength $\Gamma$. For any set of the considered parameters, in the two regions labeled D the exact solution (6) cannot exist. All the curves in Fig. 1 monotonically decrease such that for any fixed $\Gamma$ value as increasing SOC strength the maximal lattice strength associated with the exact solution has to decrease. Comparing Fig. 1(a) with 1(b), we can see that increasing the average atomic number $N_t$ per well can enlarge the region area existing exact solution for a set of fixed parameters.

\begin{center}
\includegraphics[height=1.3in,width=1.6in]{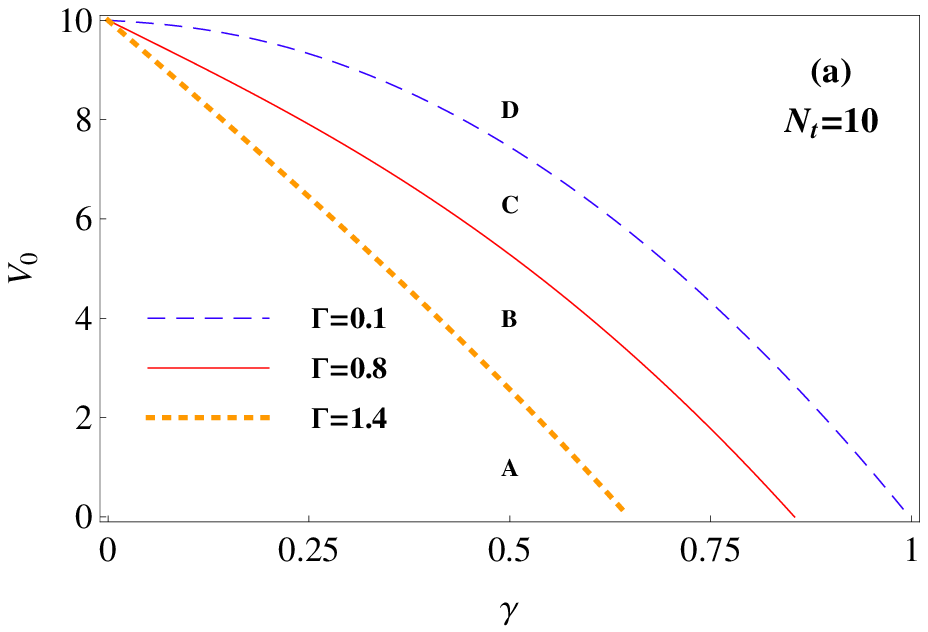}
\includegraphics[height=1.3in,width=1.6in]{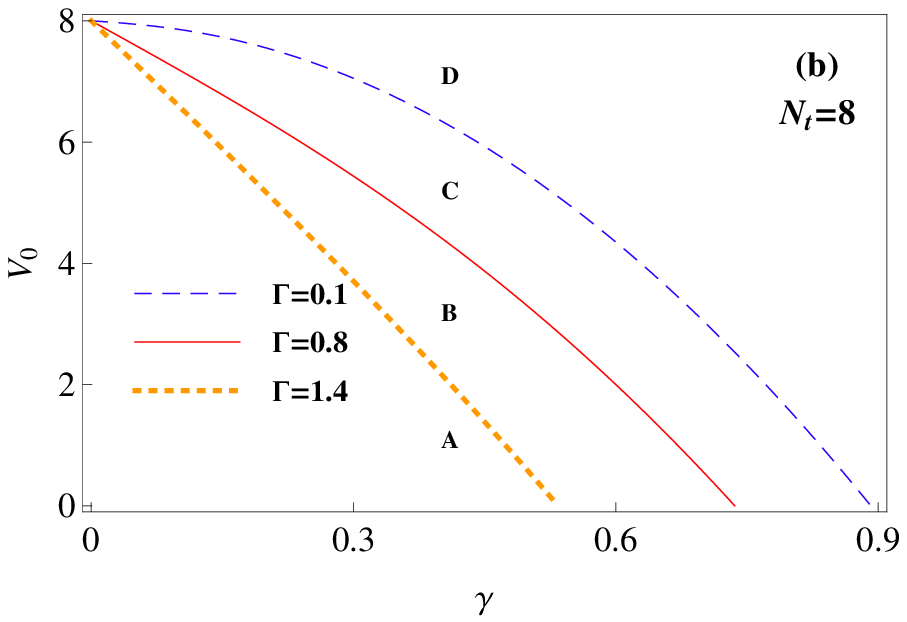}
\parbox[c]{8.0cm}{\footnotesize{\bf Fig.~1.} (color online) Lattice depth $V_{0}$ versus SOC strength $\gamma$ for the parameters $\Gamma=0.1$ (dashed curve), $\Gamma=0.8$ (solid curve) and $\Gamma=1.4$ (dotted curve), and (a) $N_{t}=10,\ g=0.6,\ g_{12}=0.4$, and (b) $N_{t}=8,\ g=0.4,\ g_{12}=0.6$. The parameter plane is divided into different regions by the boundary curves, which are labeled by A, B, C and D. Hereafter, all the quantities plotted in the figures are dimensionless.}
\end{center}

\begin{center}
\includegraphics[height=1.3in,width=1.6in]{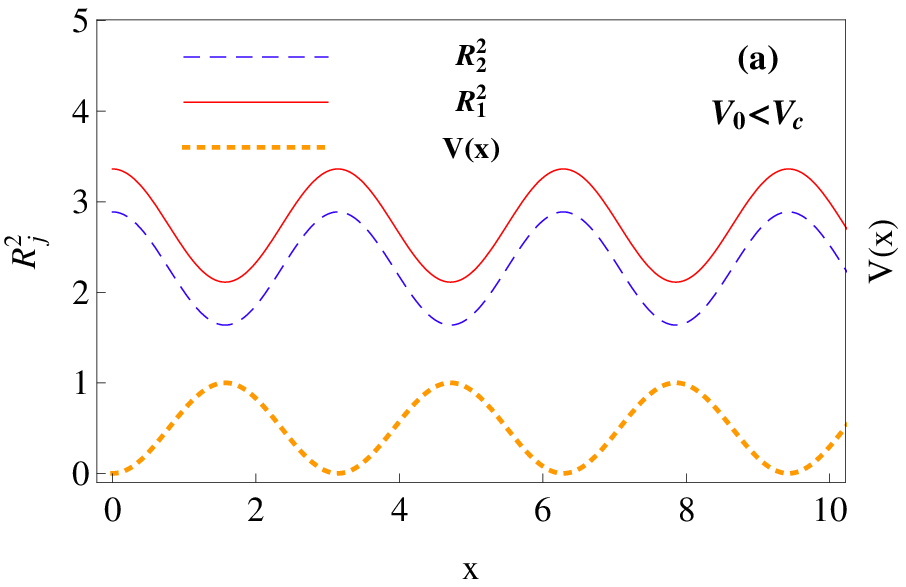}
\includegraphics[height=1.3in,width=1.6in]{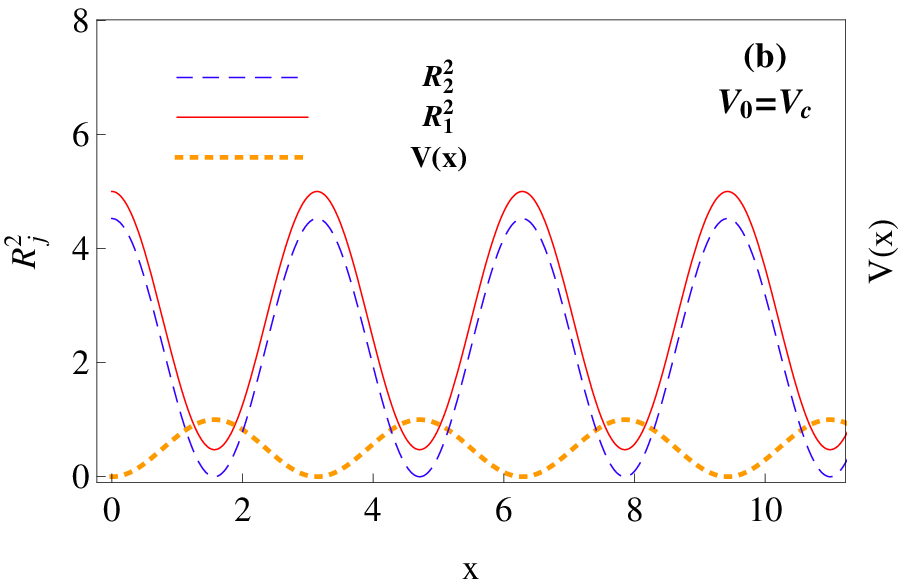}
\parbox[c]{8.0cm}{\footnotesize{\bf Fig.~2.} (color online) Spatial distributions of the atomic number density and potential function for (a) $V_{0}=1$ obeying $V_0<V_{c}$, (b) $V_{0}=3.62053$ corresponding to $V_0=V_{c}$. The atomic number density exhibits different periodic distributions without zero point in (a) and with zero points in (b). Both the figures show that every peak of the number densities aligns to a potential well.}
\end{center}

From Eq. (9) we show the spatial distributions of the atomic number density $R^{2}_{j}(x)$ and the lattice potential $V(x)$ for the parameters $g=0.6$, $g_{12}=0.2$, $\gamma=0.3$, $N_{t}=5$, $\Gamma=0.1$, and $V_{0}=1<V_c$ in Fig. 2(a), $V_{0}=3.62053=V_c$ in Fig. 2(b).
When $V_{0}< V_{c}$ is set, Fig. 2(a) displays that although the atomic number densities have different spatial distributions, their maxima align always to the center points of the potential wells, and the atomic number density is greater than zero at any spatial position. This means that more atoms are distributed around the centres of the potential wells, corresponding to higher stability of the BEC system compared to that of the case where the density peaks align to potential barrier sites and exists some zero density points.
In the case $V_{0}=V_c$, Fig. 2(b) just exhibits that $R^{2}_{j}$ maybe zero at the peak sites of potential.

\section{Physics related to the spatiotemporal Bloch states}

By applying the existence conditions of the considered exact solutions to Eq. (4) and the corresponding parameter regions, we have established the exact particular solutions (6) and (7), and illustrated the associated atomic number densities. In this section we will demonstrate that the obtained states are just the spatiotemporal Bloch states and reveal some physical properties of the BEC system in such states.

\emph{Demonstrating the spatiotemporal Bloch states}. At first from Eq. (6) we rewrite the exact solutions as $\psi_{j}=F_j(x)e^{-i x},\ F_j(x)=\frac 1 2[a_{j}(1+e^{i 2 x})-b_{j}(1-e^{i 2 x})]$ for $j=1,2$. Clearly, these are two Bloch solutions to Eq. (4) with the Bloch wave vector $-1$ and Bloch state functions $F_j(x)$ which possess the same period with the potential $V(x)$.
Then from the macroscopic quantum state $|\Psi(t)\rangle=|\Psi_{1}(t)\rangle|\uparrow\rangle+|\Psi_{2}(t)\rangle|\downarrow\rangle$ and Eq. (3) we have the space-time dependent state
\begin{eqnarray}
|\Psi(x,t)\rangle&=&\langle x|\Psi(t)\rangle=|F(x,t)\rangle e^{-i\mu t}, \nonumber\\
|F(x,t)\rangle&=&[F_{1}(x)|\uparrow\rangle+F_{2}(x)|\downarrow\rangle]e^{-ix[1+\frac{2\xi}{\omega}\sin^{2}(\frac{\omega t}{2})]}\ \ \ \
\end{eqnarray}
in the coordinate representation. Obviously, Eq. (12) is a Floquet solution to Eq. (1) with the Floquet quasienergy $-\mu$ and Floquet state $|F(x,t)\rangle$ of the same period with the driving force. At any fixed time, Eq. (12) is an usual Bloch state. Because the considered system is a time-space periodic system, we call Eq. (12) the spatiotemporal Bloch state \cite{43}. The motional state functions $F_j(x)$ contain the constants $a_j$ and $b_j$ with some arbitraries, so varying these constants can produce different spatiotemporal Bloch states.

\emph{SOC leads to generation of spin-motion entanglement}. Quantum entanglement is a universal but very special kind of quantum state in the multi-particle system. In previous work, for a SO-coupled BEC trapped in a superlattice, Kong et al. have studied the spatially chaoticity-dependent spin-motion entanglement \cite{39}. Here we study the effects of SOC on the chaoticity-independent spin-motion entanglement between internal hyperfine (pseudospin) states and external motional (orbit) states. From Eq. (7) we observe that the constants in exact solution (6) to the quasistationary Eq. (4) obey $a_1=a_2$ and $b_1=b_2$ only if $\gamma=0$. Consequently, the two motional states in Eq. (12) becomes the same, $F_1(x)=F_2(x)$, such that the quantum state $|\Psi(x,t)\rangle$ can be separated as the direct product between the spin part and motional part and become a unentangled state. The nonzero SOC strength make the different and linearly independent motional states in Eq. (12) and leads to the spin-motion entanglement.
The orbital part of the spin-motion entangled state can be used to manipulate the qubits for quantum information processing with spins.

\emph{Population imbalance of the two-component BEC}. Given Eq. (10), the difference between the average atomic numbers reads $N_{1}-N_{2}=\frac{2\gamma\sqrt{\Gamma^{2}+\gamma^{2}}}{g-g_{12}}$, which is called the population imbalance of the two-component BEC. The zero SOC strength $\gamma=0$ results in balance with the same average number per well $N_1=N_2$ and causes the entanglement loss. In the presence of SOC. for $g-g_{12}>0$, we have $N_{1}> N_{2}$ so Eq. (10) implies that only $N_2$ may reach zero with a SOC strength large enough. On the contrary, the case $g-g_{12}<0$ means that only $N_1$ may reach zero.
In Fig. 3, based on Eq. (10), we take Rabi coupling strength $\Gamma=0.1$ and average atomic number $N_t=5$ to plot the population imbalance $N_1-N_2$ versus SOC strength $\gamma$ for $g=0.6$, $g_{12}=0.2$ (solid curve), and $g=0.2$, $g_{12}=0.6$ (dotted curve). Clearly, the population imbalance monotonically increases or decreases with the increase in SOC strength for the case $g-g_{12}>0$ or the case $g-g_{12}<0$, respectively. When the SOC strength arrives at $\gamma\approx0.99750313$, we obtain $N_{1}-N_{2}=N_t=5$ or $N_{1}-N_{2}=-N_t=-5$ associated with $N_2=0$ or $N_1=0$.
Thus all the atoms are completely populated on one of the hyperfine states, as shown in Fig. 3. The results reveal the population transfer between two BEC components by adjusting the SOC strength.

\begin{center}
\includegraphics[height=1.5in,width=2.4in]{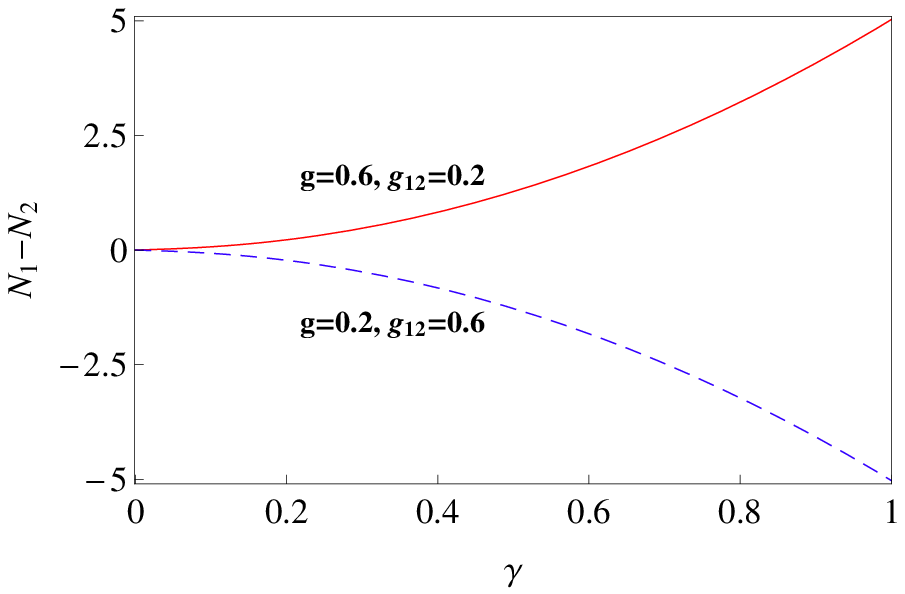}
\parbox[c]{8.0cm}{\footnotesize{\bf Fig.~3.} (color online) Plot of the population imbalance $N_1-N_2$ as a function of the SOC strength $\gamma$ for the parameters $\Gamma=0.1, N_t=5$, and $g=0.6, g_{12}=0.2$ (solid curve), and $g=0.2, g_{12}=0.6$ (dashed curve).}
\end{center}

\emph{Establishing the stable superfluid density}. Applying Eqs. (7) and (9) to the definitions of the superfluid velocity and density, $v_{j}(x)=\frac{k\hbar\Theta_{jx}}{m}=\frac{k\hbar a_{j}b_{j}}{mR_{j}^{2}}$, $J_{j}=R_{j}^{2}v_{j}=\frac{k\hbar a_{j}b_{j}}{m}$, after adopting our dimensionless parameters we get the superfluid densities and flow velocities
\begin{eqnarray}
J_{j\pm}&=&\pm\sqrt{\Big[\frac{N_{t}}{2}-(-1)^j\frac{\gamma\sqrt{\Gamma^{2}+\gamma^{2}}}{g-g_{12}}\Big]^{2}-\frac{V_{0}^{2}}{4(g+g_{12})^{2}}},\nonumber\\
v_{j\pm}(x)&=&\frac{J_{j\pm}}{R_{j}^2(x)}.
\end{eqnarray}
The constant superfluid densities describe two uniform atom currents for the two BEC components, and can be conveniently manipulated by tuning any one of the experimental parameters $N_{t}$, $g_{12}$, $g$, $V_{0}$ and recombined $\gamma$. The signs of $J_j$ determines the directions of the atomic currents, which can be initially set \cite{35,36,37} when the two species of atoms are loaded in the optical lattices and provide the persistent incident currents \cite{44}.
In Fig. 4, we plot the flow density $J_{j}$ versus the SOC strength for the parameters $\Gamma=0.1$, $N_{t}=5$, $V_{0}=1$, $g=0.6$ and $g_{12}=0.2$. When $J_j$ take the positive sign of Eq. (13), the two arrays of atomic quasi-clusters flow toward the right; when $J_{j}$ take the negative sign, the atomic quasi-clusters of the two arrays flow to the left. However, when $J_{1}$ and $J_{2}$ have different signs, the two arrays of atomic quasi-clusters flow with opposite directions. As shown in Fig. 4, the flow density $J_{1+}$ increases with increasing SOC strength $\gamma$, while $J_{2+}$ decreases with the increase in SOC strength. Conversely, $J_{j-}$ have different increasing dirrections with $J_{j+}$ for $j=1,2$. Interestingly, when we change the SOC strength to $\gamma=0.86314347$, only the superfluid density $J_1$ exists and the second component vanishes, $J_{2\pm}=0$.
From the known research results \cite{49}, when the atomic number density $R_{j}^{2}$ does not contain a zero, the BEC system may be stable. In contrast, when $R_{j}^{2}$ has a zero point, the system may be unstable for the given parameters \cite{22,26}. Such a stability is revealed from a different viewing angle by Eq. (13) where the zero density $R_{j}^{2}$ is associated with the infinite superfluid velocity resulting in the instability. Therefore, we can establish the stable superfluid currents by avoiding the critical parameter point $V_0=V_c$ and adopting the parameters in region obeying $V_0<V_c$.

\begin{center}
\includegraphics[height=1.5in,width=2.4in]{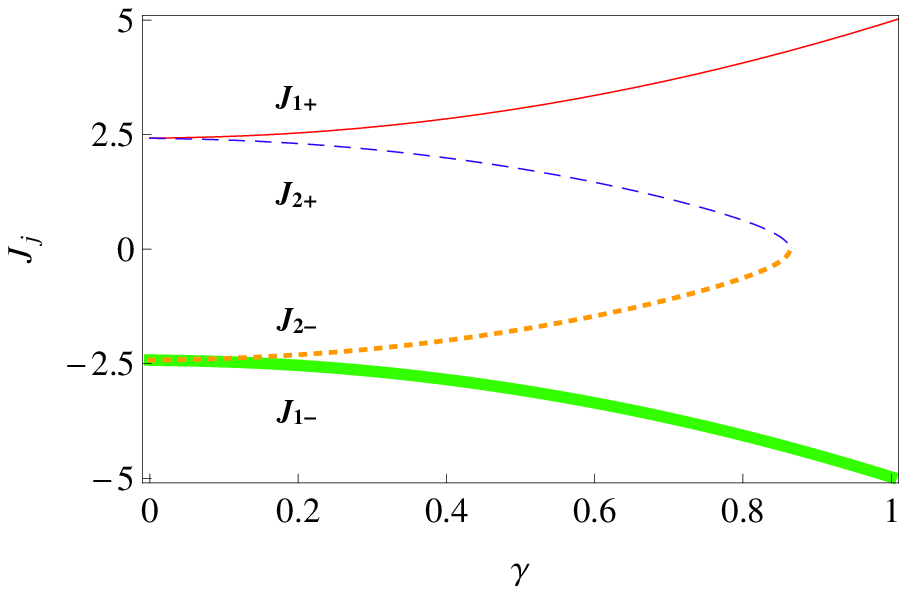}
\parbox[c]{8.0cm}{\footnotesize{\bf Fig.~4.} (color online) Plot of the flow density $J_{j}$ versus SOC strength $\gamma$. The parameters are chosen as the same as those of Fig. 2(a). The superfluid density $J_{1+}$ ($J_{2+}$) increases (decreases) with increasing SOC strength, and conversely for $J_{1-}$ and $J_{2-}$.}
\end{center}

In addition, under RWA, we can use the exact solution (6) and the corresponding parameter conditions to come up with some other interesting physical phenomena and properties.

\section{Conclusion and discussion}

In summary, we have researched the spatiotemporal Bloch states and the relating physical properties for a SO-coupled BEC trapped in an optical lattice and driven by a high-frequency field.  Based on the widely used RWA method, we obtain the exact solution (6) to the quasistationary equation (4) and the analytical spatiotemporal Bloch state (12). We establish a different method for tuning SOC via external field such that the existence conditions of the exact particular solutions are fitted. From the existence conditions we find that the parameter region on the plane of lattice- vs SOC- strength can be adjusted by the Rabi coupling strength. As the Rabi coupling strength decreases, the region areas of the parameters existing exact solution increase. Then, we show the periodic distributions of the atomic number density which are associated with the stability and instability of the system. When $V_{0}<V_{c}$ is set, more atoms are distributed around the centres of the potential wells without zero point of the atomic number density, corresponding to higher stability of the BEC system. And when $V_{0}=V_{c}$ is selected, the atomic number density contains zero points corresponding to the instability. Particularly, several new physical features related to the exact states are revealed as follows: (a) The SOC leads to generation of the spin-motion entanglement in the obtained spatiotemporal Bloch states. On the contrary, zero SOC strength results in disappearance of the spin-motion entanglement. The orbital part of the spin-motion entangled state can be used to manipulate the spin qubits for quantum information processing. (b) Tuning SOC changes the population imbalance between two BEC components and may cause what all the atoms completely concentrate on one of the hyperfine states. (c) The SOC can be applied to manipulate the stable atomic flow. For a set of fixed parameters, one component of the superfluid density decreases with the increase in SOC strength, and another component consequently increases.

BEC are well-controlled collection of atoms that can be used to study new aspects of quantum optics, many-body physics and superfluidity. What is highly desirable is to transport BEC from a stable source to a designated place \cite{44}. The results of this paper provide a method for avoiding instability of the system and for controlling BEC motion via the modulated optical lattice and the tunable SOC, particularly, are conducive to control quantum transport of BEC for different application purposes \cite{44,45}.

\end{document}